\documentclass[12pt]{article}

\usepackage{amsmath,amssymb,graphicx} 
\usepackage{epsf,color}
\usepackage{cite}

\newcommand{\drawsquare}[2]{\hbox{%
\rule{#2pt}{#1pt}\hskip-#2pt
\rule{#1pt}{#2pt}\hskip-#1pt
\rule[#1pt]{#1pt}{#2pt}}\rule[#1pt]{#2pt}{#2pt}\hskip-#2pt
\rule{#2pt}{#1pt}}

\def\wtilde{\widetilde}

\newcommand{\fund}{\drawsquare{6.5}{0.4}}
\newcommand{\afund}{\overline{\fund}}

\newcommand{\beq}{\begin{eqnarray}}
\newcommand{\eeq}{\end{eqnarray}}
\newcommand{\centeron}[2]{{\setbox0=\hbox{#1}\setbox1=\hbox{#2}\ifdim

\wd1>\wd0\kern.5\wd1\kern-.5\wd0\fi
\copy0

\kern-.5\wd0\kern-.5\wd1\copy1\ifdim\wd0>\wd1
                                       \kern.5\wd0\kern-.5\wd1\fi}}
\newcommand{\ltap}{\>\centeron{\raise.35ex\hbox{$<$}}
                               {\lower.65ex\hbox{$\sim$}}\>}
\newcommand{\gtap}{\>\centeron{\raise.35ex\hbox{$>$}}
                               {\lower.65ex\hbox{$\sim$}}\>}

\newcommand\ZZ{\hbox{\zfont Z\kern-.4emZ}}
\font\zfont = cmss10 

\begin{document}
\begin{flushright}
\end{flushright}

\vskip.5cm
\begin{center}
{\huge \bf A Simple Model of Low-scale Direct Gauge Mediation \\}

\vskip.1cm
\end{center}
\vskip0.2cm

\begin{center}
{\bf Csaba Cs\'aki$^{a}$,
Yuri Shirman$^{b}$, {\rm and}
John Terning$^{c}$}
\end{center}
\vskip 8pt

\begin{center}
$^{a}$ {\it Institute for High Energy Phenomenology\\
Newman Laboratory of Elementary Particle Physics\\
Cornell University, Ithaca, NY 14853, USA } \\
$^{b}$ {\it Department of Physics, University of California, Irvine,
CA
92697.} \\

$^{c}$ {\it
Department of Physics, University of California, Davis, CA
95616.} \\
\vspace*{0.3cm}
{\tt   csaki@lepp.cornell.edu, yshirman@uci.edu, terning@physics.ucdavis.edu}
\end{center}

\vglue 0.3truecm

\begin{abstract}
\vskip 3pt \noindent We construct a calculable model of low-energy
direct gauge mediation making use of the metastable supersymmetry
breaking vacua recently discovered by Intriligator, Seiberg and
Shih.  The standard model gauge group is a subgroup of the global
symmetries of the SUSY breaking sector and messengers play an
essential role in dynamical SUSY breaking: they are composites of a
confining gauge theory, and the holomorphic scalar messenger mass
appears as a consequence of the confining dynamics. The SUSY
breaking scale is around 100 TeV nevertheless the model is
calculable. The minimal non-renormalizable coupling of the Higgs to
the DSB sector leads in a simple way to a $\mu$-term, while the
$B$-term  arises at two-loop order resulting in a moderately large
$\tan \beta$. A novel feature of this class of models is that some particles from the 
dynamical
SUSY breaking sector may be accessible at the LHC.

\end{abstract}

\setcounter{equation}{0}
\setcounter{footnote}{0}

While supersymmetry elegantly solves the fine tuning problem of the
Higgs mass, and may even explain the origin of the weak scale by
relating it to the supersymmetry breaking scale, a generic
supersymmetric extension of the standard model (SM) itself raises a
number of problems. These problems include the $\mu$-problem (why
the single dimensionful supersymmetric parameter is related to the
supersymmetry breaking scale) and the little hierarchy problem
(which is a percent level fine-tuning problem emerging from the
non-observation of the Higgs and superpartners at LEP2).

One of the main issues that was appreciated early on in
supersymmetric model building is the problem of flavor changing
neutral currents (FCNCs): for generic soft supersymmetry breaking
scalar masses there are additional one-loop diagrams without GIM
suppression contributing to FCNC's. This problem is quite generic in
models with high scale supersymmetry breaking, where non-trivial
flavor physics is likely to affect the soft breaking scalar masses.
This issue led to interest in gauge mediated SUSY breaking~\cite{earlygmsb,gmsb} (GMSB), where
the scale of supersymmetry breaking can be below the flavor breaking
scale, and the soft masses themselves are generated via SM gauge
interactions. As a result the soft breaking mass terms will only
depend on the SM quantum numbers and be flavor independent. While
many realistic models were constructed  (see
\cite{gmsbreview} for a review), they were quite complicated and
typically had several layers of interactions (messengers) to
communicate SUSY breaking to the Standard Model fields. Simplifying
these models so that messengers would directly participate in the
dynamics of the dynamical supersymmetry breaking (DSB) sector proved
difficult. Even though viable direct gauge mediation models exist
\cite{direct}, they typically require rather large messenger scales.
While these scales could be sufficiently low to provide significant
theoretical control in studying the dynamics of the DSB sector, one
of the main promises of gauge mediation --- the possibility that in
models with a low SUSY breaking scale the DSB sector itself could in
principle be directly observable in future experiments --- was never
realized. DSB models without a hierarchy of scales are typically
strongly coupled and as a result one can at best establish the
existence of a SUSY breaking minimum but not the details of the
spectrum.

In this paper we make use of the recent discovery by  Intriligator,
Seiberg, and Shih \cite{IntriligatorSeibergShih} (ISS) of metastable
SUSY breaking vacua. From the model building point of view the main
new feature of the models of \cite{IntriligatorSeibergShih} is that
the supersymmetry breaking vacua are located near the origin of the
moduli space yet are calculable. This raises the hope that a
calculabe low scale direct mediation model can be obtained. In this
paper we show the first example of such a model. As in the ISS case
the DSB sector of our model has a fairly simple dual description in
the UV: it is just SUSY QCD with some masses and higher dimensional
operators added. The higher dimensional operators can be suppressed
by scales as high as $10^{11}$ GeV. Supersymmetry breaking is
triggered by dynamical symmetry breaking and while the SUSY breaking
scale is as low as 100 TeV, the effective low energy theory is
calculable. Since the fine-tuning depends logarithmically on SUSY
breaking scale, it is significantly reduced in our model.
Furthermore, the $\mu$-term could be generated by the same dynamics that
leads to SUSY breaking.

\vspace*{0.5cm}
 {\large \bf A simple ISS-type model}
\vspace*{0.5cm}

We will start out with a simple toy model and gradually add a few features 
in order to make a realistic
SUSY breaking model. We intend to make use of the ISS models by embedding the 
SM gauge group into the flavor symmetry of the DSB sector. Thus the flavor
symmetry should at least be SU(5). However since at least one field charged 
under the flavor symmetry gets a VEV, the minimal size of the flavor symmetry
in the DSB sector is SU(6). To focus on the simplest possibility we assume
that there is no gauge group in the magnetic description. Thus we are led to
consider the following fields charged under the global symmetries
\beq
\begin{array}{c|ccc}
 & SU(6)& U(1) & U(1)_R\\ \hline
\tilde \phi & \fund &   1 & 0 \vphantom{\sqrt{{\wtilde M}}}\\
{\bar {\tilde \phi} }& \overline{\fund}   & -1 & 0 \vphantom{\sqrt{{\wtilde M}}}\\
{\wtilde M} & {\bf Ad}+{\bf 1} & 0 & 2 \\
\end{array}~,
\eeq with the superpotential \beq \label{wone} W_1 ={\bar {\tilde
\phi} }{\wtilde M}{\tilde  \phi} -h f^2 {\rm Tr}  {\wtilde M} ~.
\eeq The global symmetries of this model are just the symmetries of
an s-confining $SU(5)$ gauge  theory with 6 massive flavors
\cite{Seiberg:1994bz-9,CsakiSchmaltzSkiba}. Indeed, we can identify
$\wtilde M$, $\tilde\phi$, and $\phi$ with mesons, baryons, and
antibaryons of the electric description respectively. The linear
term in the superpotential above then arises from a mass term in the
microscopic theory with the identification $m \Lambda \approx h
f^2$, while the cubic term is required to ensure the correct mapping
of the two descriptions. Finally  instantons generate an  operator
\beq \label{wdyn}
W_{\rm inst.}=\frac{\det \widetilde M}{\Lambda^3},
\eeq
where
$\Lambda$ is the intrinsic  holomorphic (dynamical) scale of  the microscopic theory.
As explained in ref.~\cite{IntriligatorSeibergShih}, the term in Eq.~(\ref{wdyn}) is
repsonsible for ensuring that there is a SUSY preserving global
minimum with \beq \langle {\wtilde M} \rangle \sim  f
\left(\frac{\Lambda}{f}\right)^\frac{3}{5}~. \eeq There is also a
metastable SUSY breaking vacuum at $\langle {\wtilde M} \rangle
\sim0$, which can have a lifetime much longer than the age of the
Universe for $f \ll \Lambda$. Near this SUSY breaking vacuum, the
instanton term is an irrelevant operator that we can ignore (unless
we want to calculate the tunnelling rate to the true vacuum state).
SUSY is broken since the matrix ${\bar {\tilde \phi} }{\tilde  \phi}
$ has rank one, so \beq \frac{\partial W_1}{\partial  {\wtilde
M}^J_I } = {\bar {\tilde \phi} }^I{\tilde  \phi}_J- h f^2 \delta_J^I
\ne 0~. \eeq In order minimize the scalar potential energy, one
flavor (in an appropriate basis) will get a VEV,  ${\bar {\tilde
\phi} }^K{\tilde  \phi}_K= h f^2 $, and the global $SU(6)$ symmetry
will be spontaneously broken to $SU(5)$.

\vspace*{0.5cm}
   {\large \bf Gauging the flavor symmetry}
\vspace*{0.5cm}

We will gauge the standard model (SM) subgroup of $SU(5)$ in which
case vacuum alignment \cite{vacuumalignment} will prefer the VEV to
align so as to preserve the gauge symmetry.  Thus it is convenient to
write the fields in a form where the unbroken symmetry is manifest,
so we split $\tilde \phi_J$ into $\phi_j$ and $\psi$ (where $\langle
\psi \rangle \ne0$) and similarily \beq {\wtilde M}=\left(
\begin{array}{cc}
 M^j_i & N^j \\
 {\bar N}_i & X \end{array}\right)~.
\eeq
So our field content can be rewritten as~\footnote{We only show quantum
numbers under symmetries relevant for the following analysis.}

\beq
\label{fields}
\begin{array}{c|cccccccc}
 & \phi  &\bar\phi &\psi&\bar\psi & M & X & N & \bar N\\ \hline
 SU(5) \vphantom{\wtilde M}& \fund & \afund &  1 & 1 & {\bf Ad}+{\bf 1} & 1 & \fund & \afund \\
U(1)_R \vphantom{\wtilde M}& 0  &0  & 0 & 0 & 2 & 2 & 2 & 2\\
\end{array}~,
\eeq with a superpotential \beq W_2 ={\bar \phi }  M \phi +{\bar
\psi} X \psi+ {\bar \phi} N \psi+{\bar \psi} {\bar N} \phi -h f^2
\left( {\rm Tr}    M +X\right)~. \label{sp} \eeq

Let us discuss the dynamics entailed by this superpotential. The
equation of motion for $X$ leads to a non-zero $\bar \psi \psi$ VEV.
This in turn marries $\phi~(\bar \phi)$ with $\bar N~(N)$ making
sure that they are massive and do not obtain VEVs. Finally, SUSY is
broken by the ${\mathcal F}$-component of ${\rm Tr} M$. Rescaling
$\mathrm{Tr}M$ so  that its kinetic term is canonically normalized,
we obtain \beq {\mathcal F}_{{\rm Tr} M} = \sqrt{5} h f^2~. \eeq As
pointed out in \cite{IntriligatorSeibergShih} there are a number  of
massless states at  tree-level. Some of these are goldstone bosons
of the spontaneously broken $SU(6)$ symmetry. Since $SU(6)$ can be
explicitly broken by the superpotential and is certainly broken by
gauging the SM subgroup, these fields can obtain masses at one-loop
level once SUSY is broken. There are also massless scalars
corresponding to pseudo-flat directions of the O'Rafeartaigh model.
As shown in \cite{IntriligatorSeibergShih} these will also obtain
soft masses at one-loop through which can be easily analyzed using
the
 Coleman-Weinberg potential. In
particular, the field with the non-vanishing ${\mathcal F}$-term,
${\rm Tr} M$, is stabilized at the origin. Finally, the fermionic
components of $M$ remain massless.

\vspace*{0.5cm}
   {\large \bf Communicating SUSY breaking to the SM fields}
\vspace*{0.5cm}

Let us now describe how SUSY breaking is communicated to  the SM
superpartners. The fields $\bar \phi$ and $\phi$ couple directly to
the SUSY breaking and  obtain holomorphic soft mass terms. In
addition, $\bar N$ and $N$ will obtain holomorphic soft mass terms
due to the supersymmetric mixing (through $\langle \psi \rangle$)
with $\bar\phi$ and $\phi$ in the superpotential. Once the SM
subgroup of $SU(5)$ is gauged these fields will act as messengers.
It is important to notice that in the absence of a VEV for ${\rm Tr}
M$ the mass matrix of scalar messengers would have one vanishing
eigenvalue at tree level. This is problematic for two reasons.
First, we cannot allow light scalars with SM charges. Second,
messenger multiplets with a vanishing supertrace of the mass matrix
and massless scalars do not usually contribute to soft scalar masses
at two-loops (although non-trivial mixing of the messengers in our
model modifies the calculation of gauge mediated contributions).
There is one more problem in the model presented so far -- at the
minimum of the potential there is an accidental discrete
$R$-symmetry which forbids gaugino
masses~\cite{IntriligatorSeibergShih}. In this case this is a
$Z_{10}$ R-symmetry. This is the discrete subgroup of the U(1)$_R$
in (\ref{fields}) that is left unbroken by (\ref{wdyn}). In fact,
the same $R$-symmetry forbids the soft masses for fermionic
components of $M$. Thus in order to solve these problems we need to
generate a scalar VEV for the field $M$. Below we will show two
possibilities for how to achieve that.

Once we have a VEV  for ${\rm Tr} M$, the fermion messenger mass
matrix will have the see-saw form \beq m_f=\left(
\begin{array}{cc}
 \langle {\rm M}\rangle& \langle \psi \rangle\\
\langle \bar\psi \rangle & 0\\
\end{array}
\right )~. \eeq

A Majorana gaugino mass will now be generated at one-loop as in more
standard gauge mediation models. Notice that to leading order in the
SUSY breaking parameter ${\mathcal F}$ the gaugino mass is
proportional to ${\rm Tr} (m_f^{-1} {\mathcal F})$ and it vanishes
in our model.\footnote{We thank Hitoshi Murayama for pointing this
out to us.}
 However, terms which are higher order in ${\mathcal F}$
are non-vanishing and we have verified that the ${\mathcal
F}^3/m_f^5$ term is non-vanishing. Since ${\mathcal F}/m_{f}^2 \sim
1$ in our model these contributions are not suppressed. The masses
of both the scalar and fermionic superpartners  (while qualitatively
similar) will differ from predictions of usual gauge mediation
scenarios, due to a more complicated messenger spectrum. 

Scalar components of $M$ will obtain contributions to their masses
both  from the Coleman-Weinberg potential and the usual gauge
mediated loops (except for two scalars that are neutral under the
Standard Model). Gauge mediated contributions will be dominant for
sufficiently small $h$ expected from an underlying microscopic
description. Finally, we note that the fermionic components of $M$ will
obtain masses both from gauge mediation and from Yukawa coupling to
messengers, $\bar \phi M \phi$. Thus these particles may have TeV scale masses, and 
 be accessible at the LHC. Since $M$ contains $SU(5)$ adjoints some of the scalars
 and fermions will be colored and thus can be  easily produced. The details of the spectrum
 and experimental signatures
 will be studied elsewhere.

\vspace*{0.5cm}
   {\large \bf Generating the $M$-VEV via singlet interactions}
\vspace*{0.5cm}

We have seen above that in order to generate a non-vanishing gaugino
mass we need to generate a scalar VEV for $M$. Here and in the next
paragraph we show two possibilities for that. While the models seem
quite similar the implications of the two mechanisms are actually
quite different. In both cases we modify the theory slightly by
adding the singlets $S,\bar{S}, Z, \bar{Z}$. In the first case these
fields will be fundamental singlets and the discrete R-symmetry is broken
via explicit superpotential interactions:
\beq W &=&{\bar \phi } M
\phi +{\bar \psi} X \psi+ {\bar \phi} N \psi+{\bar \psi} {\bar N}
\phi-hf^2\left( {\rm Tr}   M +X\right) \nonumber \\
&&+(d \,{\rm Tr} M +m)S {\bar S} + m^\prime(S\bar Z + Z \bar S)~.
\label{suppot} \eeq The new superpotential terms necessarily break
the $R$-symmetry, due to the simultaneous presence of both the ${\rm
Tr} M\,S\bar{S}$ and the $S\bar{S}$ terms. The last terms in this
superpotential (proportional to $m'$) ensure that supersymmetry is
not restored via VEV's of $S$ and $\bar S$. There are now two
interactions contributing to the one-loop potential for $M$. Loops
of the singlets $S$ and $\bar{S}$ will tend to generate an $M$ VEV
to cancel the mass term for these fields while loops of $\bar \phi$
and $\phi$ will generate a positive contribution to the mass squared
of ${\rm Tr} M$. Generically the vacuum will be shifted from the
origin. In order to keep the $S,\bar{S}$ fields from obtaining VEV's
(so that SUSY breaking originates fully from the F-term of $M$, and 
not partly from the singlet sector)
we need to assume $m'^2> d h f^2$. The one-loop Coleman-Weinberg
potential for spontaneously broken supersymmetric theories
\begin{equation}
\frac{1}{64\pi^2} {\rm STr} {\cal M}^4 \log \frac{{\cal
M}^2}{\Lambda^2}
\end{equation} can be evaluated for the potential following from
(\ref{suppot}). In order for the minimum of the potential to be
significantly shifted from the origin, the interaction strength of
the $S,\bar{S}$ fields to ${\rm Tr}M$ should not be very small,
otherwise their effect around the origin will be negligible. As an
example we show the Coleman-Weinberg potential along the ${\rm Tr}M$
direction in units of $\sqrt{h f^2}$, for parameters $m=4, m'=1.5,
d=0.4$ (again in the same units). We can see that for these
parameters the minimum is at $\mathrm{Tr} M\sim  \sqrt{5 hf^2}$.
This implies that messenger multiplets  obtain supersymmetric
contributions to their masses, $m_m$, which are comparable to
splittings withing the multiplet, ${\cal F}/m_m^2 \sim 1$.

\begin{figure}[tb]
\begin{center}
\includegraphics[width=5cm]{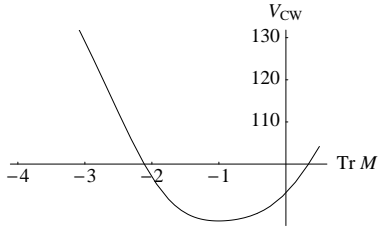}
\label{CW}
\end{center}
\caption{Plot of the Coleman-Weinberg  potential (in arbitrary
units). The horizontal axis is ${\rm Tr}M$ in units of $\sqrt{{\cal
F}}$.}
\end{figure}

\vspace*{0.5cm}
   {\large \bf Generating the $M$-VEV via gauge interactions}
\vspace*{0.5cm}

A perhaps more elegant solution for generating the VEV for $M$ is by
using the mechanism of~\cite{Dine}. Instead of adding the explicit
$mS\bar{S}$ mass term one can maintain the discrete R-symmetry in
the superpotential, and only break it spontaneously via the VEV of
$M$. This has the added benefit that imposing this discrete
R-symmetry can forbid some (but not all) other unwanted terms in the superpotential
(for example ${\rm Tr}M^2$ which would restore supersymmetry). 

To achieve the spontaneous breaking of the R-symmetry (by forcing
the $M$ VEV from the origin) we   gauge a U(1) symmetry
under which $S,Z$ have charges +1 and $\bar{S},\bar{Z}$ have charges
$-1$. The superpotential will now be \beq W &=& {\bar \phi } M \phi
+{\bar \psi} X \psi+ {\bar \phi} N \psi+{\bar \psi} {\bar N}
\phi-hf^2\left( {\rm Tr}   M +X\right)  \nonumber \\ &+&d \,{\rm Tr}
M S {\bar S} + m^\prime(S\bar Z + Z \bar S)~. \label{suppot2} \eeq

\begin{figure}[tb]
\begin{center}
\includegraphics[width=5cm]{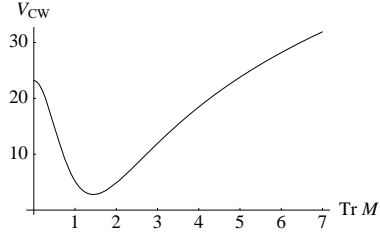}
\end{center}
\caption{Plot of the Coleman-Weinberg  potential for the case with a U(1) gauge symmetry (in arbitrary
units). The horizontal axis is ${\rm Tr}M$ in units of $\sqrt{{\cal
F}}$.}
\label{fig:CW2}
\end{figure}

In order for these U(1) gauge fields to contribute to the CW
potential we need to pick the parameters of the theory such that $S$
obtains a VEV. The reason for the additional contributions to the CW
potential is that in this case the $Z,\bar{Z}$ directions will be
related to the $M$ VEV. As a consequence SUSY breaking will not fully 
originate from the dynamical sector, but the $S,Z$ sector will also contribute. 
Minimizing the CW potential one can find minima similar to those
in~\cite{Dine} for a wide range of perturbative U(1) couplings $g$
and small couplings $d$. An example for such a minimum is shown in
Fig.~\ref{fig:CW2} for $d=.01$, $g=0.1$, and $m^\prime=0.05$ 
(in units of $\sqrt{{\mathcal F}}$).

\vspace*{0.5cm}
   {\large \bf Generating the SUSY breaking linear term via dynamics}
\vspace*{0.5cm}

We now turn to the origin of the linear term which leads to SUSY
breaking. It can be generated from a condensate of an additional
supercolor sector (this possibility has recently been also
emphasized in~\cite{DFS}). One of the simplest possibilities is an
$SU(2)_{sc}$ gauge group with 2 flavors. Thus our complete SUSY
breaking model has the following fields \beq
\begin{array}{c|ccccccccccccccc}
           & \phi &\bar \phi & \psi & \bar \psi & M & X & N & \bar N & S & \bar S & Z & \bar Z & p & \bar p & T\\ \hline
SU(2)_{sc} \vphantom{\wtilde M}& 1    &   1      &    1 & 1         & 1 & 1 & 1 &   1    & 1 & 1      & 1 &    1   & \fund & \afund & 1\\ 
U(1)_{gauge}    \vphantom{\wtilde M} & 0    &  0       &   0  &   0       & 0 & 0 & 0 & 0 & 1   & -1 &  1     & -1 &   0    & 0 &   0    \\
\hline
SU(5)    \vphantom{\wtilde M}  & \fund & \afund  & 1    &    1      & {\bf Ad+1} & 1 & \fund & \afund & 1 & 1 & 1 & 1 & 1 & 1   & 1\\
SU(2) & 1 &1&1&1&1&1&1&1&1&1&1&1& \fund &\afund &1 \\
\end{array}~.
\label{charges} \eeq

The full superpotential of the model is
\beq
W &=&{\bar \phi }  M \phi +{\bar \psi} X \psi+ {\bar \phi} N \psi+{\bar \psi} {\bar N} \phi-\frac{h}{\Lambda^2_{UV}}p^2 {\bar p}^2\left( {\rm Tr}   M +X\right) \nonumber \\
&&+c \, T   \,p {\bar p}+d \,{\rm Tr} M S {\bar S} +
m^\prime(S\bar Z + Z \bar S)~. \label{fullsuppot} \eeq

This superpotential can be enforced for example by the discrete $Z_{10}$ R-symmetry under which
$M,X,N,\bar{N}, Z, \bar{Z}, T$ have charge 2, and $p,\bar{p},\phi ,\bar{\phi}, \psi ,\bar{\psi}, S, \bar{S}$ 
have charge 0. This is anomaly free under the SU(2)$_{sc}$ and the 
U(1)$_{gauge}$. However, it is still not the most generic superpotential term allowed by the symmetries. 
The term $XS\bar{S}$ would also be allowed by the symmetries, but has to be assumed to  vanish.
However, other dangerous terms like ${\rm Tr}\ M^2$ are excluded by this discrete symmetry.
The couplings to the singlet $T$  give mass terms for the
mesons $p {\bar p}$. Since the supercolor sector has a deformed
quantum moduli space this leaves the ``baryons" $B=p^2$ and $B={\bar
p}^2$, with non-zero VEVs.  Thus the strong dynamics enforces \beq
p^2 {\bar p}^2 =B{\bar B}=\Lambda_{sc}^4= f^2 \Lambda_{UV}^2 \eeq
where $\Lambda_{sc}$ is the intrinsic holomorphic scale for the
supercolor gauge group, and the last part of the equation can be taken as the definition of the scale $f$.

\vspace*{0.5cm}
   {\large \bf Generating the $\mu$-term}
\vspace*{0.5cm}

To complete the construction of a phenomenologically relevant model
we need  to generate the terms in the  Higgs potential necessary for
electroweak symmetry breaking.
 To obtain a $\mu$ term
we add the superpotential term
\beq W _\mu= \beta \, \frac{p^2 {\bar
p}^2}{\Lambda_{UV}^3} \, H_u  H_d~.
\label{muterm}
\eeq
Then around the scale where
the supercolor group gets strong the model generates a $\mu$ term of order
\beq \mu \sim \beta\, f\left( \frac{\Lambda_{sc}}{\Lambda_{UV}}
\right)^2~.
\eeq
The soft SUSY-breaking $B$-term, however, vanishes at tree level. As
noticed in \cite{variations} a $B$-term is however generated at two
loops (since $M_2$ itself is generated at one-loop order)
\beq
B \sim \frac{3\, \alpha_2}{2\pi} M_2 \mu\, \ln \left( \frac{\cal
F}{M_2 \mu} \right)\, .
\eeq
This leads to a $B$-term that is small
compared to the square of electroweak scale (by a factor of
$\alpha_2$) and consequently results in a large $\tan\beta$, of order 10--50.

\vspace*{0.5cm}
   {\large \bf The microscopic description of the theory}
\vspace*{0.5cm}

We will assume that all the
singlets $S$, $\bar S$, $Z$, $\bar Z$, and $T$ are elementary fields both in the effective and microscopic
description.
Above the scale $\Lambda$ the microscopic  dual  description is a generalized SUSY QCD:
\beq
\begin{array}{cccc|cccc}
&SU(5) & SU(2)_{SC} & U(1)_{gauge} & SU(5)&SU(2)&SU(2)& Z_{10} \\ \hline
Q& \fund & {\bf 1}& 0 &\fund & {\bf 1}& {\bf 1} &  1 \\
{\bar Q}& \overline{\fund}  & {\bf 1}& 0 & \overline{\fund}   & {\bf 1}& {\bf 1} & 1\\
q& \fund & {\bf 1}& 0 & {\bf 1} & {\bf 1}& {\bf 1} &   1 \\
{\bar q}& \overline{\fund}  & {\bf 1}& 0 & {\bf 1}   & {\bf 1}& {\bf 1} & 1 \\
p& {\bf 1}& \fund & 0 & {\bf 1} &\fund & {\bf 1}  &0\\
{\bar p}& {\bf 1}& \fund & 0 & {\bf 1}& {\bf 1}  &\fund &0\\
S& {\bf 1}& {\bf 1}& 1 & {\bf 1}& {\bf 1}& {\bf 1} &0 \\
{\bar S}& {\bf 1}& {\bf 1}& -1 & {\bf 1}& {\bf 1}& {\bf 1} &0 \\
Z& {\bf 1}& {\bf 1}& 1 & {\bf 1}& {\bf 1}& {\bf 1} &2 \\
{\bar Z}& {\bf 1}& {\bf 1}& -1 & {\bf 1}& {\bf 1}& {\bf 1} &2 \\
T& {\bf 1}& {\bf 1}& 0 & {\bf 1}&\fund&\fund &2 \\
\end{array}~,
\label{electric}
\eeq
with a superpotential
\beq
W &=&m \,S {\bar S}  + m^\prime(S\bar Z + Z \bar S)+c \, T   \,p {\bar p}  +\beta \, \frac{p^2 {\bar
p}^2}{\Lambda_{UV}^3} \, H_u  H_d \\ \nonumber &&+\frac{\tilde d}{\Lambda_{UV}} \,{\rm Tr} \,Q{\bar Q} \,S {\bar S} -\frac{\tilde h}{\Lambda^3_{UV}}p^2 {\bar p}^2\left( {\rm Tr} \,Q{\bar Q}+q{\bar q} \right) ~.
\label{finalmodel}
\eeq

The mapping between the two descriptions is  $M \leftrightarrow Q \bar Q /\Lambda$,   $X
\leftrightarrow q \bar q /\Lambda$,  $N \leftrightarrow Q \bar q
/\Lambda$, etc. Thus  (with $\tilde h$ and $\tilde d$ of order 1)
we expect that natural values of $h$ and $d$ are of order
\beq h
\sim d\sim \frac{\Lambda}{\Lambda_{UV}}~.
 \eeq

\vspace*{0.5cm}
   {\large \bf Estimate of scales}
\vspace*{0.5cm}

We are now ready to present some typical  values for the scales in this
theory. To have a low-scale model of SUSY breaking we are assuming that
$\sqrt{{\cal F}} \sim 100$ TeV. We are also assuming that the messenger mass scale
determined by Tr $M$ is of the same order $\sim 100$ TeV. The highest possible value
for the UV scale $\Lambda_{UV}$ can be obtained by calculating the
Landau pole for the QCD coupling. This depends quite sensitively on the
mass of the components of the SU(5) adjoint field $M$. If we assume
that the leading contribution to their masses (along with the superpartners
of the SM fields) are the gauge mediated contributions at 1 TeV, then
we find the scale for the Landau pole to be around few$\cdot 10^{11}$ GeV.
This can be increased slightly (by a factor of at most 10) by decreasing the ratio
$\Lambda /\Lambda_{UV}$. Thus a safe assumption would be a choice
satisfying $\Lambda_{UV}\leq 10^{11}$ GeV. Finally, we need to generate a $\mu$-term
of order 100 GeV. The scales thus should satisfy the relations:
\begin{equation}
\frac{\Lambda \Lambda_{sc}^4}{\Lambda_{UV}^3}\sim {\cal F} \sim
(100\ {\rm TeV})^2, \ \ \frac{\Lambda_{sc}^4}{\Lambda_{UV}^3} \sim
\mu \sim  100\ {\rm GeV},\ \Lambda_{UV} \leq 10^{11}
\ {\rm GeV}.
\end{equation}
Assuming we choose the parameters corresponding to the potential in
Fig.~\ref{fig:CW2} we can satisfy these constraints by:
\begin{eqnarray}
&& \Lambda \sim \Lambda_{sc} \sim 10^8 \ {\rm GeV}, \ \Lambda_{UV} \sim 10^{10} \ {\rm GeV},
\ \ m' \sim 5 \cdot 10^3 {\rm GeV}, \ \tilde{d} \sim 1.
\end{eqnarray}
In this case the bounce action interpolating between the SUSY breaking vacuum and the SUSY preserving vacuum
can be estimated to be
\begin{equation}
S_{b} \sim (\Delta M )^4/{\cal F}^2 \sim 5 \cdot 10^8,
\end{equation}
thus the tunneling rate is suppressed by a factor of $e^{-10^8}$, which suggest that the metastable vacuum
should have a lifetime $\tau$ much longer than the age of the Universe ($\sim 4 \times 10^{17} s$).
A back-the-envelope-estimate gives
\beq
\tau \sim \frac{1}{100\, {\rm GeV}} \, \frac{1 s}{10^{24} \,{\rm GeV}^{-1}}\,\sqrt{ \frac{2 \pi}{S_{b}}} \, e^{S_{b}}
\sim 10^{2 \cdot 10^8} s~.\eeq

\vspace*{0.5cm}
   {\large \bf The little hierarchy}
\vspace*{0.5cm}

 The minimal model described above provides a simple implementation of minimal gauge
mediation with a single messenger field (up to the deviations
discussed in the previous section due to the mixing of the messenger
field with non-messengers). One of the main drawbacks of minimal
gauge mediation is the little hierarchy problem. There are usually three separate
sources for the little hierarchy problem in gauge mediated models. The first and most important
is the the large mass ratio of the squark and slepton/Higgs masses dictated by the quantum numbers
of the messenger fields, and is specific to gauge mediated models. The other two sources of fine tuning
are generic to supersymmetric extensions of the SM.
One of these is the logarithmic running of the soft mass parameters, which
in this case is cut off at the messenger scale and could logarithmically enhance the soft mass parameters
appearing in the Higgs potential, which  thus usually requires a small stop mass to avoid fine-tuning.
Finally, one also needs to make sure that the Higgs mass is above the
115 GeV LEP2 bound, which usually requires a heavy stop mass.

One of these issues is naturally resolved here, since we can take
the messenger scale to be around 100 TeV, therefore the logarithmic
enhancement of the soft masses is very small. The stop/slepton mass
ratio can be lowered for example by changing the number of doublet
messengers vs. triplet messengers~\cite{Kaustubh}. Another even
simpler possibility would be to change the hypercharge assignments
of the messengers. However, these are usually incompatible with
unification. Unification is however problematic in our model anyway
due to the Landau pole.  At the scale $\Lambda_{UV}$ one would need
to UV complete the model, which would likely involve taking a dual
of the color gauge group, resulting in a cascading gauge theory,
without a conventional perturbative unification (but rather unifying
onto string theory in a warped throat~\cite{KS}).

  As for the Higgs bound, it strongly
depends on how the $\mu$-term is  generated. With the operator given
in Eq. (\ref{muterm}) there is no additional Higgs quartic term
generated, however, one can easily imagine extensions of this model
where the operator leading to the $\mu$-term will contain NMSSM-type
additional sources for a quartic term thus relaxing the fine-tuning
from the Higgs mass constraint.

\vspace*{0.5cm}
   {\large \bf Summary}
\vspace*{0.5cm}

We have presented a calculable low-scale model of direct gauge
mediation. The supersymmetry breaking scale can be as low as 100
TeV, and there is no hierarchy between the messenger masses and the
SUSY breaking scale. The messengers play an integral part in SUSY
breaking: they are composites of the dynamics that breaks
supersymmetry and in the magnetic picture it is the structure of the
interactions of the messengers that actually results in SUSY
breaking. These interactions are such that the SUSY breaking
holomorphic mass term for the messengers emerges naturally. In order
to generate a real mass for the messengers one needs to shift the
VEV of the meson $M$ from the origin, which can be achieved by
adding additional interaction terms in the superpotential or by
including an additional U(1) gauge interaction. A $\mu$-term
can be generated from dynamics, and some  of the sources of the SUSY
fine tuning problem can be eliminated. We expect that the phenomenology of the model will be quite distinctive
due to the presence of additional TeV scale particles and modifications of the traditional GMSB spectrum.
The main drawback for now is
the usual Landau-pole problem that is simply appearing to the the
presence of the SU(5) adjoint superfield $M$.

One of the most interesting features of the model is a rich phenomenology at the 
TeV scale. Particles from the DSB sector may be directly observable at the LHC. 
In particular,  both the scalar and fermionic components of 
$M$ can have TeV scale masses. 
Scalars in $M$ obtain mass due to the Coleman-Weinberg potential 
as well as gauge mediated contributions at two-loops. While the Coleman-Weinberg 
potential is generated at one-loop, its contribution to scalar masses scales 
as $h^4 f^2$ and is small in a large part  of the parameter space. Therefore we expect the scalar 
masses to be dominated by GMSB contributions. Fermionic 
components of $M$ obtain mass both from GMSB loops as well as one-loop 
contributions arising from $M \phi \bar \phi$ coupling. The latter coupling is of 
order one, so the new fermions will roughly have masses comparable to the gluino mass. 
The new scalar and fermionic particles at the TeV scale include $SU(5)$ adjoints which transform as $(8,1)_0$, $(1,3)_0$, $(3, 
2)_{1/6}$, and $(\bar 3, 2)_{-1/6}$ under $SU(3)_c \times SU(2)_w\times 
U(1)$.

\section*{Acknowledgements}
We thank  Kaustubh Agashe, Zackaria Chacko, Spencer Chang, Michael
Dine, Hitoshi Murayama and David Shih for useful discussions. We
thank the Aspen Center for Physics where this work was initiated.
The research of C.C. is supported in part by the DOE OJI grant
DE-FG02-01ER41206 and in part by the NSF grant PHY-0355005. J.T. is
supported by the US Department of Energy grant DE-FG02-91ER40674.

\section*{Note Added}
While this manuscript was in preparation
Refs.~\cite{KitanoHiroshi,MN} appeared, which also make use of
metastable vacua for gauge mediation. In Refs.~\cite{MN} the
messengers do not play an essential role in the supersymmetry breaking dynamics.
The main difference from Ref.~\cite{KitanoHiroshi} is that there
the R-symmetry is broken via a mass term $N\bar{N}$ (in the
notation of Eq.(\ref{sp})) while here this is achieved by generating a VEV
for $M$. A preliminary version of this work has been presented in~\cite{Yuritalk}.


\begin{thebibliography}{99}

\bibitem{earlygmsb}
  M.~Dine, W.~Fischler and M.~Srednicki,
  Nucl.\ Phys.\ B {\bf 189}, 575 (1981);
S.~Dimopoulos and S.~Raby,
  Nucl.\ Phys.\ B {\bf 192}, 353 (1981);
 M.~Dine and W.~Fischler,
  Phys.\ Lett.\ B {\bf 110}, 227 (1982);
M.~Dine and M.~Srednicki,
  Nucl.\ Phys.\ B {\bf 202}, 238 (1982);
L.~Alvarez-Gaume, M.~Claudson and M.~B.~Wise,
  Nucl.\ Phys.\ B {\bf 207}, 96 (1982);
C.~R.~Nappi and B.~A.~Ovrut,
  Phys.\ Lett.\ B {\bf 113}, 175 (1982).


\bibitem{gmsb}
  M.~Dine and A.~E.~Nelson,
  Phys.\ Rev.\ D {\bf 48}, 1277 (1993)
  [arXiv:hep-ph/9303230];
  M.~Dine, A.~E.~Nelson and Y.~Shirman,
  Phys.\ Rev.\ D {\bf 51}, 1362 (1995)
  [arXiv:hep-ph/9408384];
 M.~Dine, A.~E.~Nelson, Y.~Nir and Y.~Shirman,
  Phys.\ Rev.\ D {\bf 53}, 2658 (1996)
  [arXiv:hep-ph/9507378].

\bibitem{gmsbreview}
G.~F.~Giudice and R.~Rattazzi,
  Phys.\ Rept.\  {\bf 322}, 419 (1999)
  [arXiv:hep-ph/9801271].

\bibitem{direct}
  E.~Poppitz and S.~P.~Trivedi,
  Phys.\ Rev.\ D {\bf 55}, 5508 (1997)
  [arXiv:hep-ph/9609529];
 N.~Arkani-Hamed, J.~March-Russell and H.~Murayama,
  Nucl.\ Phys.\ B {\bf 509}, 3 (1998)
  [arXiv:hep-ph/9701286];
  H.~Murayama,
  Phys.\ Rev.\ Lett.\  {\bf 79}, 18 (1997)
  [arXiv:hep-ph/9705271];
  K.~I.~Izawa, Y.~Nomura, K.~Tobe and T.~Yanagida,
  Phys.\ Rev.\ D {\bf 56}, 2886 (1997)
  [arXiv:hep-ph/9705228];
  M.~A.~Luty,
  Phys.\ Lett.\ B {\bf 414}, 71 (1997)
  [arXiv:hep-ph/9706554];
  S.~Dimopoulos, G.~R.~Dvali and R.~Rattazzi,
  Phys.\ Lett.\ B {\bf 413}, 336 (1997)
  [arXiv:hep-ph/9707537];
  Y.~Shirman,
  Phys.\ Lett.\ B {\bf 417}, 281 (1998)
  [arXiv:hep-ph/9709383];
  K.~Agashe,
  Phys.\ Lett.\ B {\bf 435}, 83 (1998)
  [arXiv:hep-ph/9804450].

\bibitem{IntriligatorSeibergShih}
  K.~Intriligator, N.~Seiberg and D.~Shih,
  JHEP {\bf 0604} (2006) 021
  [arXiv:hep-th/0602239].

\bibitem{Seiberg:1994bz-9}
N.~Seiberg,
  Phys.\ Rev.\ D {\bf 49}, 6857 (1994)
  [arXiv:hep-th/9402044].

\bibitem{CsakiSchmaltzSkiba}
 C.~Cs\'aki, M.~Schmaltz and W.~Skiba,
  Phys.\ Rev.\ Lett.\  {\bf 78}, 799 (1997)
  [arXiv:hep-th/9610139];
Phys.\ Rev.\ D {\bf 55}, 7840 (1997)
  [arXiv:hep-th/9612207].


\bibitem{vacuumalignment}
  S.~Weinberg,
  Phys.\ Rev.\ D {\bf 13} (1976) 974.
   S.~Weinberg,
  Phys.\ Rev.\ D {\bf 19} (1979) 1277.
  M.~E.~Peskin,
  Nucl.\ Phys.\ B {\bf 175} (1980) 197.
  J.~Preskill,
  Nucl.\ Phys.\ B {\bf 177} (1981) 21.

\bibitem{Dine}
M.~Dine and J.~Mason,
  arXiv:hep-ph/0611312.

\bibitem{DFS}
  M.~Dine, J.~L.~Feng and E.~Silverstein,
  arXiv:hep-th/0608159.

\bibitem{variations}
  M.~Dine, Y.~Nir and Y.~Shirman,
  Phys.\ Rev.\ D {\bf 55}, 1501 (1997)
  [arXiv:hep-ph/9607397].

\bibitem{Kaustubh}
  K.~Agashe and M.~Graesser,
  Nucl.\ Phys.\ B {\bf 507}, 3 (1997)
  [arXiv:hep-ph/9704206].

\bibitem{KS}
I.~R.~Klebanov and M.~J.~Strassler,
  JHEP {\bf 0008}, 052 (2000)
  [arXiv:hep-th/0007191].


\bibitem{KitanoHiroshi}
R.~Kitano, H.~Ooguri and Y.~Ookouchi,
  arXiv:hep-ph/0612139.

\bibitem{MN}
H.~Murayama and Y.~Nomura,
  arXiv:hep-ph/0612186.

\bibitem{Yuritalk}
Talk presented by Y.~Shirman at UC Davis (11/06/06) and at UC Berkeley (11/13/06),
{\tt http://particle.physics.ucdavis.edu/seminars/data/media/2006/nov/shirman.pdf}

\end{thebibliography}
\end{document}